# Analysis of theoretical NMR spectra generated by exact solutions of the Bloch-McConnell and the Bloch-Torrey equations for a two-compartment radial diffusive exchange model

Mihai R. Gherase

Physics Department, Mount Allison University, 67 York Street, Sackville, New Brunswick, E4L 1E6, Canada
E-mail: mgherase@mta.ca

**Abstract**

Nuclear Magnetic Resonance (NMR) spectra shaped by diffusive exchange processes are encountered in biological and medical applications. Two models based on the Bloch-McConnell (B-M) and the Bloch-Torrey (B-T) equations are commonly used for modelling the physical processes which determine the NMR lineshapes. Qualitative arguments for each of the two methods can be found in various studies in the literature. However, there is a lack of systematic quantitative investigations of the diffusive exchange spectra calculated with the two methods for the same physical system or model. In this work exact frequency-domain transverse magnetization solutions of the B-M and the B-T equations with boundary conditions for a two-compartment radial diffusive exchange model are presented. The average lifetimes in each compartment ($\tau_1$ and $\tau_2$) were calculated using the pre-exchange lifetimes method at equilibrium. The broadening of the theoretical spectra due to the diffusive exchange process was analyzed using two metrics which were calculated numerically: the Full Width at Half Maximum (FWHM) and the Full Width at Tenth Maximum (FWTM). Theoretical spectra and the two corresponding metrics were computed by varying three different parameters: diffusive permeability of the separating membrane between the two compartments ($P$), the radius of the inner spherical compartment ($a$), and the chemical shift between the two compartments ($\Delta f$). For simplicity, the diffusion coefficients of spin-bearing particles in the two compartments were set constant ($D_1 = D_2 = 5 \times 10^{-9}$ m$^2$ s$^{-1}$) and the volumes of the two compartments were set to be equal ($V_1 = V_2$). In general, the two models predicted different spectral broadening for the same parameters in agreement with the different analytical solutions. However, the numerical analysis of the spectral broadening demonstrated that the two models converge to the same results in the limit of large chemical shift on the scale of the exchange rate ($\Delta f / \tau_e^{-1} > 1$). The results were qualitatively interpreted based on the difference between the two models: the B-M model assumes a single average exchange time constant while the B-T model implies a more realistic continuous distribution of diffusive exchange times.

## 1. Introduction

In the NMR literature the diffusive exchange term refers to the diffusion-mediated motion of spin-bearing nuclei between two or multiple compartments. In this context, the term compartment designates a microscopic sample volume characterized by constant transverse and longitudinal relaxation times ($T_1$ and $T_2$), and constant Larmor resonant frequency ($\omega_0$). Alternatively, the term *site* is also used, in most cases indicating a volume of molecular or atomic scale. The chemical exchange is a closely related term which describes essentially the same process: the back-and-forth motion of nuclear spins between two or more sites. The different terminology only denotes the underlying process of the spin kinetics: chemical reaction or diffusion. The first study of chemical exchange on the observed NMR spectrum was reported in 1956 by Gutowsky and Holm [1]. Since that time, many chemical exchange studies have been reported. Two review articles are found in references [2] and [3].

Diffusive exchange processes are particularly important in NMR studies of the morphology and function of biological structures. Diffusive exchange of atoms and molecules with spin-bearing nuclei between intra- and inter-cellular compartments is a transverse and longitudinal relaxation mechanism which inherently shapes the NMR data [4]-[8]. Data analysis of these NMR experiments relies on diffusive exchange models. Two diffusive exchange models are prevalent in the existent literature. They are based on two different sets of equations which were derived from the phenomenological Bloch equations [9] by adding a term on their right-hand side. These are the Bloch-McConnell [10] and the Bloch-Torrey equations [11]. The added term in the Bloch-McConnell equations accounts for a constant exchange rate of the spins between the two compartments, while the additional term in the Bloch-Torrey equations incorporates the diffusion of spins. Therefore, the areas of applicability are different: the Bloch-McConnell equations can be applied to chemical exchange processes, while the Bloch-Torrey equations are suitable for modeling diffusive exchange systems.

Exact and approximate solutions of the Bloch-McConnell equations were derived by Leigh and his collaborators [12]-[14]. Although the exact solutions appear lengthy, it was demonstrated that, in certain situations, they are necessary for accurate analysis and interpretation of experimental data [15]. For NMR applications involving two-compartment diffusive exchange systems, the solutions of Bloch-McConnell equations require the knowledge of the spin average lifetimes in each compartment at equilibrium: $\tau_1$ and $\tau_2$. The average lifetimes can be calculated using diffusion theory and compartment geometry for two-compartment systems with a non-permeable separating membrane [16]-[18]. For two-compartment systems with a separating permeable membrane the solution was indicated by Chen and Springer [19] and used in later studies [6]-[8].

Belton and Hills employed the Laplace transform method to solve the transverse magnetization part of the Bloch-Torrey equations in the frequency domain [20], [21]. They derived analytical expressions of the NMR spectra for one-dimensional, two-compartment and single-compartment with surface relaxation systems using the appropriate boundary conditions. The same approach was also employed by Gherase *et al.* [22] to derive exact analytical expressions of the transverse magnetization for a two-compartment radial diffusive exchange model with a permeable membrane. Although the initial goal of this model was to analyze the NMR free-induction decay (FID) spectra obtained from a particular physical system, its generality and finite volume geometry justifies an extended analysis.

For brevity, the approach using the Bloch-McConnell equations with lifetimes calculations will be referred from here on as the B-M model. Similarly, the Bloch-Torrey with boundary solutions approach will be referred as the B-T model. It appears there is no consensus in the literature with regard to which of the two models is best suited for describing a particular diffusive exchange system. Authors who used the B-M model in their investigations invoked the following arguments: the Bloch-McConnell equations are well-known and well-studied, they have exact solutions, and they successfully modeled the experimental data in various NMR applications with diffusive exchange [6]-[8]. On the other hand, authors who were in favour of the B-T model, claimed their approach is more realistic in describing the NMR diffusive exchange relaxation process [5], [20]-[22]. It was argued that the average lifetime method does not take into account the localization of the spins with respect to the boundary where the exchange takes place. On the scale of the transverse relaxation time constant, the diffusive exchange involves more spins closer to the boundary and less when further away. In other words, the spins within a single compartment are not "equivalent" in a diffusive exchange process as is assumed by the B-M model [5]. A distribution of exchange times is a better representation of the spin kinetics than a single average value. Although this reasoning highlights a fundamental difference between the two models, its quantitative significance for the analysis of NMR spectra is not completely elucidated and requires further investigation. A study of the differences between the theoretical NMR spectra calculated with the two models has the following advantages: (i) the diffusive exchange is *a priori* established as the only additional transverse relaxation mechanism, (ii) the results are known within the precision of the numerical methods employed without the limitations associated with experimental uncertainties, and (iii) this study can potentially improve the interpretation and analysis of experimental data.

This paper describes a comparative study between the theoretical spectra computed using the B-M and B-T models applied to a two-compartment radial diffusive exchange system [22]. The work is organized in five sections. Exact analytical solutions for the complex transverse magnetization are presented for both models in section 2 and appendices A, B, and C. Theoretical spectra were generated for a wide range of exchange times by varying two physical parameters: the radius of the inner spherical compartment ($a$) and the diffusive permeability ($P$) of the separating membrane. Computational methods used in the generation and numerical analysis of the theoretical spectra are described in section 3. Section 4 contains the results of the analysis. The physical interpretation and discussion of these results are found in section 5.

## 2. Theory

This section is organized in two sub-sections. The first sub-section contains the mathematical expressions describing the theoretical spectra obtained by taking the Fourier transform of the time-domain solutions of the Bloch-McConnell (B-M) equations with the calculated pre-exchange lifetimes. The second sub-section contains the frequency-domain analytical solutions of the Bloch-Torrey (B-T) equations with boundary solutions. Both methods were applied for a radial diffusive exchange system consisting of two concentric spherical compartments separated by a thin permeable membrane. The schematic representation is shown in Fig. 1. The physical and geometrical parameters characterizing the system are provided in Table 1.

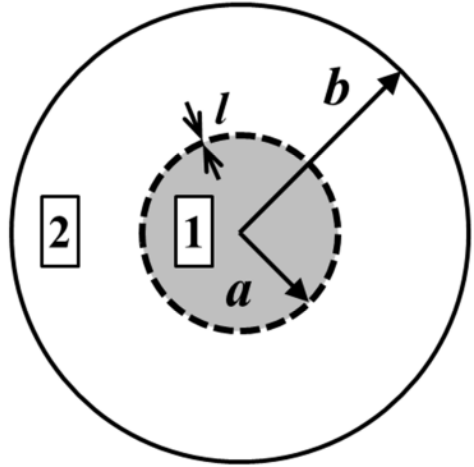

**Fig. 1**. Schematic representation of the two-compartment radial diffusive exchange system. The two compartments are denoted as 1 and 2 in the figure. The spherical compartment 1 is shown in the grey colour and its radius is $a$. The radius of the sphere enclosing the two compartments is $b$. The dotted line represents the permeable membrane separating the compartments. The membrane is considered thin ($l \ll a$).

**Table 1**
Geometrical and physical parameters used in the B-M and B-T models.

| Symbol | Description | Model | Dimension |
|---|---|---|---|
| $a$ | Radius of inner sphere (compartment 1) | B-M, B-T | [length] |
| $b$ | Radius of outer sphere (compartments 1 and 2) | B-M, B-T | [length] |
| $l$ | Membrane thickness | B-M, B-T | [length] |
| $\omega_0$ | Larmor resonant angular frequency | B-M, B-T | $[\text{time}]^{-1}$ |
| $R$ | Transverse relaxation rate constant | B-M, B-T | $[\text{time}]^{-1}$ |
| $M$ | Complex transverse magnetization | B-M, B-T | $[\text{length}]^{-3}$ |
| $P$ | Membrane permeability | B-M, B-T | $[\text{length}] \times [\text{time}]^{-1}$ |
| $D$ | Diffusion coefficient | B-T | $[\text{length}]^2 \times [\text{time}]^{-1}$ |
| $\tau$ | Average lifetime | B-M | [time] |

## 2.1. B-M model

The macroscopic Bloch-McConnell equations describing the time dependence of the transverse magnetizations in the two compartments in the presence of chemical exchange are given by:

$$\frac{dM_1}{dt} = -(R_1 - i\omega_{01})M_1 - \tau_1^{-1}M_1 + \tau_2^{-1}M_2, \tag{1}$$

$$\frac{dM_2}{dt} = -(R_2 - i\omega_{02})M_2 - \tau_2^{-1}M_2 + \tau_1^{-1}M_1. \tag{2}$$

The calculations of the average lifetimes ($\tau_1$ and $\tau_2$) are based on two equations:

$$\tau_1^{-1} = P\frac{3}{a}, \tag{3}$$

$$\frac{\tau_1}{\tau_2} = \frac{V_1}{V_2} = \left(\frac{a}{b-a}\right)^3. \tag{4}$$

The notations used in Eqs. (3)-(4) are provided in Table 1. The subscripts indicate the compartment. An important parameter is the exchange time ($\tau_e$), or exchange rate ($\tau_e^{-1}$), and is calculated using the following relationship:

$$\tau_e^{-1} = \tau_1^{-1} + \tau_2^{-1}. \tag{5}$$

The quantity $3/a$ from Eq. (3) is the area-to-volume ratio ($A/V$) for the spherical compartment 1. The relationship: $\tau^{-1} = P(A/V)$, where $A/V$ is the area-to-volume ratio, was initially derived in reference [19]. The physical significance of the average lifetimes ($\tau_1$ and $\tau_2$) and the exchange rate ($\tau_e^{-1}$) is given in Appendix A where Eqs. (3) and (4) were derived from first principles.

The initial conditions for the complex transverse magnetizations ($M_{01}$ and $M_{02}$) are given below:

$$M_{01} = |M_{01}| = \Re(M_{01}), \tag{6a}$$

$$M_{02} = |M_{02}| = \Re(M_{02}), \tag{6b}$$

$$M_{01} + M_{02} = 1. \tag{6c}$$

The notations $\Re(z)$ and $\Im(z)$ were used to denote the real and imaginary parts of complex number $z$, respectively. Equations (6a) and (6b) also imply that $\Im(M_{01}) = 0$ and $\Im(M_{02}) = 0$, or that the initial phase of the transverse magnetization was set to zero. The solutions of Eqs. (1) and (2) were taken from reference [15] and are provided in the Eqs.(7a) and (7b). The notations used are given in Eqs. (7c)-(7l).

$$M_1(t) = \Theta_1 e^{\vartheta_1 t} + \Theta_2 e^{\vartheta_2 t}, \tag{7a}$$

$$M_2(t) = \Theta_3 e^{\vartheta_1 t} + \Theta_4 e^{\vartheta_2 t}, \tag{7b}$$

where:

$$\vartheta_1 = \frac{1}{2}\left[-(k_1 + k_2) + \sqrt{(k_1 - k_2)^2 + 4(\tau_1 \tau_2)^{-1}}\right], \tag{7c}$$

$$\vartheta_2 = \frac{1}{2}\left[-(k_1 + k_2) + \sqrt{(k_1 - k_2)^2 - 4(\tau_1 \tau_2)^{-1}}\right], \tag{7d}$$

$$\Theta_1 = \frac{1}{\vartheta_1-\vartheta_2}[-(\vartheta_2+k_1)M_{01}+\tau_2^{-1}M_{02}], \tag{7e}$$

$$\Theta_2 = \frac{1}{\vartheta_1-\vartheta_2}[(\vartheta_1+k_1)M_{01}+\tau_2^{-1}M_{02}], \tag{7f}$$

$$\Theta_3 = \frac{1}{\vartheta_1-\vartheta_2}[\tau_1^{-1}M_{01}+(\vartheta_1+k_1)M_{02}], \tag{7g}$$

$$\Theta_4 = \frac{1}{\vartheta_1-\vartheta_2}[-\tau_1^{-1}M_{01}-(\vartheta_2+k_1)M_{02}], \tag{7h}$$

$$k_1 = R_1 - i\omega_{01} + \tau_1^{-1}, \tag{7i}$$

$$k_2 = R_2 - i\omega_{02} + \tau_2^{-1}. \tag{7l}$$

The sum of Eqs. (7a) and (7b) represents the time dependence of the FID corresponding to the two-compartment system:

$$\text{FID}(t) = (\Theta_1+\Theta_3)e^{\vartheta_1 t}+(\Theta_2+\Theta_4)e^{\vartheta_2 t}. \tag{8}$$

The Fourier transform of the $\text{FID}(t)$ function from Eq. (8) is denoted with $\mathcal{F}[\text{FID}(t)]$ and is calculated analytically in Appendix B. Its real part represents the spectrum $S(\omega)$ and is given in Eq. (9). The mathematical details of these calculations are also provided in Appendix B.

$$S(\omega) = \frac{-\mathfrak{R}(\Theta_1+\Theta_3)\mathfrak{R}(\vartheta_1)+\mathfrak{I}(\Theta_1+\Theta_3)(\omega-\mathfrak{I}(\vartheta_1))}{\mathfrak{R}(\vartheta_1)^2+(\omega-\mathfrak{I}(\vartheta_1))^2}+\frac{-\mathfrak{R}(\Theta_2+\Theta_4)\mathfrak{R}(\vartheta_2)+\mathfrak{I}(\Theta_2+\Theta_4)(\omega-\mathfrak{I}(\vartheta_2))}{\mathfrak{R}(\vartheta_2)^2+(\omega-\mathfrak{I}(\vartheta_2))^2}. \tag{9}$$

## 2.2. B-T model

In the Bloch-Torrey equations, transverse magnetization ($M$) depends on time ($t$) and spatial coordinates ($x, y, z$). In this model, isotropic diffusion is assumed. Therefore, in spherical coordinates ($r, \theta, \varphi$) the transverse magnetization ($M$) depends only on the radial component ($r$) and time ($t$), and is given by:

$$\frac{\partial M}{\partial t} = -i\omega_0 M - RM + D\frac{1}{r}\frac{\partial^2}{\partial r^2}(rM). \tag{10}$$

The Laplace transform of the transverse magnetization $M$ is denoted with $\widetilde{M}$ and is given by:

$$\widetilde{M}(r,s) = \int_0^\infty e^{-st} M(r,t)\, dt, \tag{11}$$

where $s = -i\omega$. Using the definition of $\widetilde{M}$ from Eq. (11) and integrating by parts, Eq. (10) becomes:

$$D\frac{1}{r}\frac{\partial^2}{\partial r^2}(\widetilde{M}r) - \alpha\widetilde{M} = -M_0. \tag{12}$$

where $\alpha = s + i\omega_0 + R$. The diffusive exchange and transverse relaxation processes are described by the following two differential equations with four additional equations representing the boundary conditions. The subscripts indicate the compartment.

$$D_1\frac{1}{r}\frac{\partial^2}{\partial r^2}(\widetilde{M_1}r) - \alpha_1\widetilde{M_1} = -M_{01},\ 0 \le r \le a, \tag{13a}$$

$$D_2\frac{1}{r}\frac{\partial^2}{\partial r^2}(\widetilde{M_2}r) - \alpha_2\widetilde{M_2} = -M_{02},\ a \le r \le b. \tag{13b}$$

$$D_1\left(\frac{\partial\widetilde{M_1}}{\partial r}\right)_{r=a} = P[\widetilde{M_2}(a)-\widetilde{M_1}(a)], \tag{14a}$$

$$D_2 \left(\frac{\partial \widetilde{M}_2}{\partial r}\right)_{r=a} = P[\widetilde{M}_2(a) - \widetilde{M}_1(a)], \tag{14b}$$

$$\left(\frac{\partial \widetilde{M}_1}{\partial r}\right)_{r=0} = 0, \tag{14c}$$

$$\left(\frac{\partial \widetilde{M}_2}{\partial r}\right)_{r=b} = 0. \tag{14d}$$

Equations (14a) and (14b) are mathematical expressions for the net flux of spins across the separating membrane, which is zero at equilibrium. Equations (14c) and (14d) express a zero net flux of spins at the boundaries of the radial diffusive system. The initial conditions are given by Eqs. (6a)-(6c) from subsection 2.1. The solution of Eqs. (13a) and (13b) with boundary conditions given by Eqs. (14a)-(14d) can be found in reference [22]. The equations were rearranged to express the final solution in a simpler form. The main steps of the mathematical derivation are provided in Appendix C. The complex quantity $S(s)$ from the following equation represents the sum of the complex magnetizations $\widetilde{M}_1(r,s)$ and $\widetilde{M}_2(r,s)$ integrated over the volumes of the two compartments: $V_1$ and $V_2$.

$$S(s) = \frac{M_{01}}{\alpha_1}(V_1 + V_{ex}) + \frac{M_{02}}{\alpha_2}(V_2 - V_{ex}), \tag{15}$$

In this equation the following notations are used:

$$\alpha_j = s + i\omega_{0j} + R_j = R_j - i(\omega - \omega_{0j}), \quad j = \{1,2\}, \tag{16a}$$

$$\xi_j = \left(\frac{\alpha_j}{D_j}\right)^{1/2}, \quad j = \{1,2\}, \tag{16b}$$

$$V_1 = \frac{4\pi a^3}{3}, \tag{16c}$$

$$V_2 = \frac{4\pi(b^3 - a^3)}{3}, \tag{16d}$$

$$V_{ex} = -\frac{4\pi a^2 P(\alpha_1^{-1} - \alpha_2^{-1})}{Pa\left(\frac{F_2}{D_2} - \frac{F_1}{D_1}\right) - 1}, \tag{16e}$$

$$F_1 = \frac{\tanh(\xi_1 a)}{\xi_1 a - \tanh(\xi_1 a)}, \tag{16f}$$

$$F_2 = \frac{\xi_2 b - \tanh[\xi_2(b-a)]}{(1 - \xi_2^2 ab)\tanh[\xi_2(b-a)] - \xi_2(b-a)}. \tag{16g}$$

The real quantities $\Re(M_{01}/\alpha_1)$ and $\Re(M_{02}/\alpha_2)$ represent the Lorentzian lines corresponding to the non-exchange case. The complex function $V_{ex}$ from Eq. (16e) has volumetric dimension and contains all the physical and geometrical parameters of the diffusive exchange process described in Table 1. The complex functions $F_1$ and $F_2$ are non-dimensional. Therefore, the denominator of complex function $V_{ex}$ is also non-dimensional. The NMR spectrum corresponding to this model $S(\omega)$ is given by:

$$S(\omega) = \frac{\Re[S(s)]}{2\pi}. \tag{17}$$

## 3. Methods

The NMR exchange spectra represented by Eqs. (9) and (17) were calculated for a given set of physical and geometrical parameters using two custom C++ codes. The spectra were generated by calculating the numerical value of $S(\omega)$ for each value of $\omega = 2\pi f$. The frequency range was calculated as a function of the chemical shift $\Delta f$ and spanned from $f_{min} = 0$ Hz to $f_{max} = 4\,\Delta f$. The resonance frequencies $f_{01}$ and $f_{02}$ were set symmetrically with respect to $f_{min}$ and $f_{max}$:

$$f_{01} = 3/2\,\Delta f \text{ and } f_{02} = 5/2\,\Delta f. \tag{18}$$

The spectrum calculations were performed in equal steps of 0.02 Hz.

The membrane permeability ($P$) was varied in the range $(10^{-2} - 10)$ µm s$^{-1}$ corresponding to water permeability in biological cells (references (6) and (23), and references therein) in 100 equal steps. Three different values were selected for the radius of the spherical compartment ($a$): 1, 5, and 10 µm. These values are on the size scale of biological cells. The intrinsic transverse relaxation rate constants in the two compartments were set to be: $R_1 = R_2 = 10$ s$^{-1}$. The diffusion coefficients values were set to be: $D_1 = D_2 = 5 \times 10^{-9}$ m$^2$ s$^{-1}$. This value is in the middle of the range for water diffusion coefficients measured in human tissues $(10^{-9} - 10^{-8})$ m$^2$ s$^{-1}$ (24). The volumes of the two compartments were set to be equal: $V_1 = V_2$. Hence, the radius $b$ of the two-compartment model depicted in Fig. 1 can be calculated as a function of radius $a$ according to:

$$b = 2^{1/3}a. \tag{19}$$

The spectral broadening was assessed using two metrics: the Full Width at Half Maximum (FWHM) and the Full Width at Tenth Maximum (FWTM) which were calculated numerically. These metrics are commonly used in the scientific literature and their definitions are well-known. However, the coalescing of the two resonant peaks with increasing exchange rate requires extended definitions of these two metrics. Figure 2 and its caption contain the extended definitions of the FWHM and FWTM for any spectra with one or two distinguishable peaks. The numerical algorithms employed in these calculations were composed of a quadratic interpolation to determine each peak position and a linear interpolation to determine the frequencies corresponding to the half-maximum and tenth-maximum amplitudes.

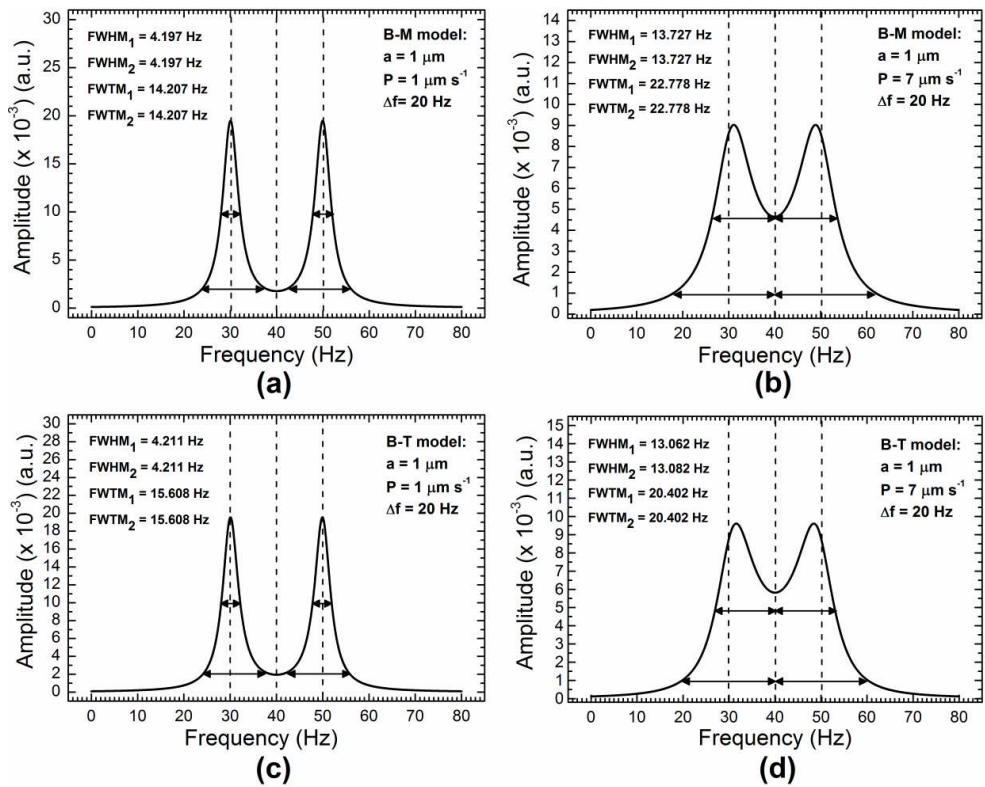

**Fig. 2.** Four panels (a)-(d) show theoretical spectra with two distinguishable peaks and the corresponding graphical representations of the FWHM and FWTM metrics for two different permeability values (P = 1 μm and P = 7 μm). Panels (a) and (b) show the spectra generated using the B-M model; panels (c) and (d) show the corresponding spectra generated using the B-T model. The vertical dotted lines indicate the two resonant frequencies and their average. The legend at the top provides the corresponding parameters and the numerically calculated FWHM and FWTM values for each peak, graphically represented by the horizontal double arrow lines. The FWHM and FWTM metrics of the entire spectrum were calculated as the sums of the two values corresponding to each peak. High diffusive exchange rate spectra with one distinguishable peak are not shown here. In that case, the two metrics were calculated following the usual definitions.

## 4. Results

Sample plots of the theoretical spectra generated using the two models are shown in Fig. 3 for a single inner radius ($a = 1$ μm), a single chemical shift value ($\Delta f = 10$ Hz), and four values of the membrane permeability ($P = 1, 2, 4, 8$ μm s$^{-1}$).

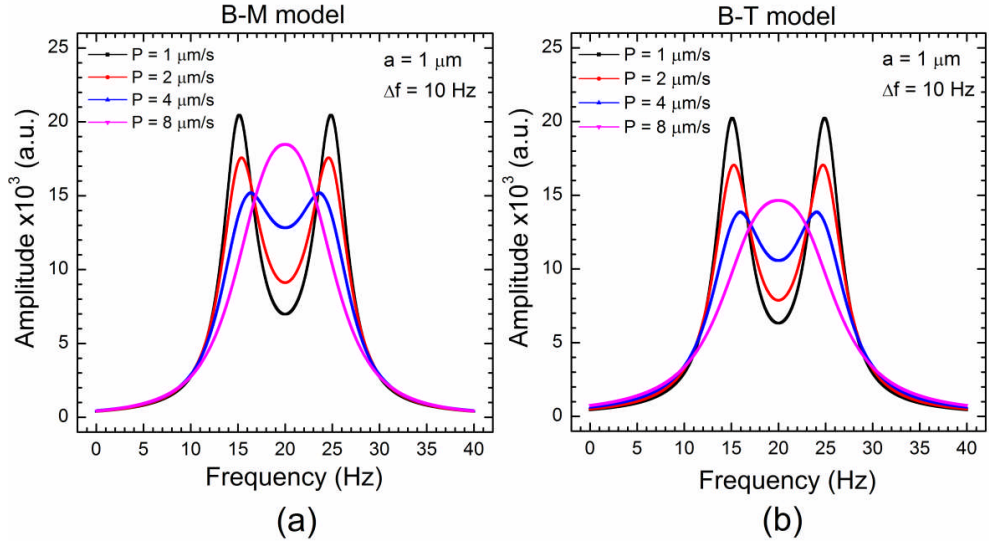

**Fig. 3.** Sample plots of theoretical diffusive exchange spectra generated using (a) the B-M model and (b) the B-T model. The resonant frequencies are 15 Hz and 25 Hz. Parameters $a$ and $\Delta f$ are given in the legend and in the upper right-hand corner of the two graphs. The step size was 0.02 Hz; hence, these two plots contain $40/0.02 = 2000$ sample points each.

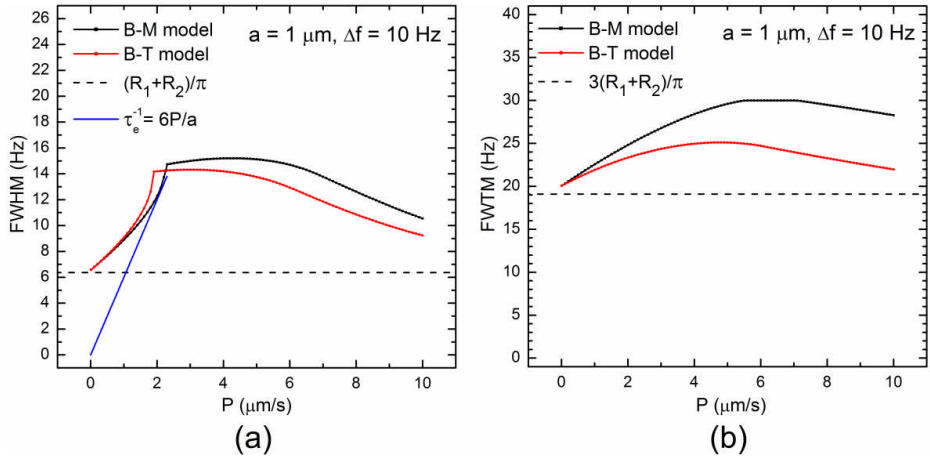

**Fig. 4.** Sample plots of numerically calculated metrics: (a) FWHM and (b) FWTM for the theoretical spectra generated using the B-M model (black) and the B-T model (red). The dashed lines represent the non-exchange limits as represented by two Lorentzian resonant peaks. $R_1$ and $R_2$ are the transverse relaxation time constants corresponding to the two compartments as indicated in Table 1. The blue line from plot (a) represents the exchange rate $\tau_e^{-1}$ calculated for membrane permeability ($P$) values between 0 and 2.4 μm/s. The intersection between the blue and black lines indicates where the approximation $FWHM \cong \tau_e^{-1}$ is strictly valid within the B-M model. The slope discontinuity in 4(a) is due to the coalescing of the two resonant peaks. The inner radius ($a$) and chemical shift ($\Delta f$) are indicated in the upper-right corner.

Figure 4 shows two plots of the FWHM (a) and FWTM (b) as a function of membrane permeability $P$ on the range: $(10^{-2} - 10)$ μm s$^{-1}$. The two dotted lines show the non-exchange limit for FWHM and FWTM of the spectra. The inflection points seen in Fig. 4(a) indicate the coalescing of the two peaks.

The quantitative differences between the theoretical spectra generated by the two models are shown in Fig. 5 for three different values of inner radius $a$. It can be seen that the FWTM ratio (FWTM$_{BM}$/FWTM$_{BT}$) is larger than the FWHM ratio (FWHM$_{BM}$/FWHM$_{BT}$) for the same physical parameters. Also, it can be seen that the plots corresponding to larger diameters ($a = 5$ μm and $a = 10$ μm) are similar scaled-down versions of the plot corresponding to $a = 1$ μm. For higher values of radius $a$, there is a better agreement between the two models for the same range of membrane permeability $P$.

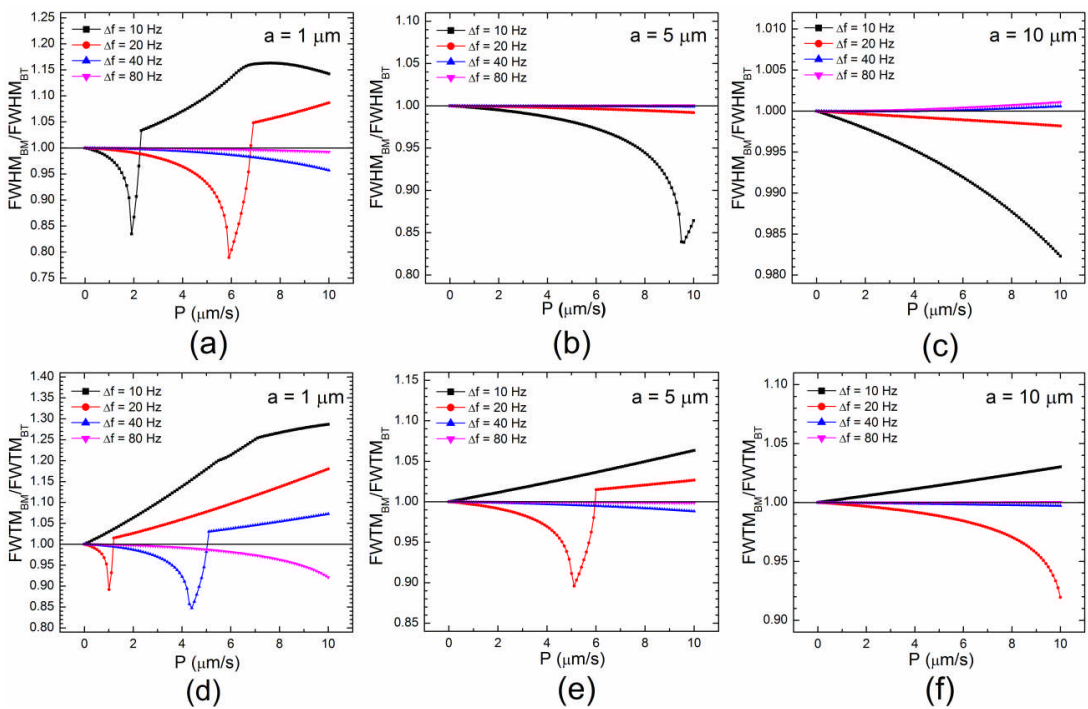

**Fig. 5.** The upper-half panels (a)-(c) show plots of the ratio between the FWHM of the theoretical spectra generated using the B-M model and that of spectra generated using the B-T model (FWHM$_{BM}$/FWHM$_{BT}$) for three different radii ($a$) and four different chemical shifts ($\Delta f$). The lower-half panels (d)-(f) show the corresponding plots of the FWTM ratios. The horizontal continuous lines indicate the equality between the metrics corresponding to the two models.

The coalescing of the two resonant spectra for various diffusive exchange regimes (i.e. varying $P$) can be observed in the plots of Fig. 6. The FWTM/FWHM ratios corresponding to the B-M and B-T models are plotted as a function of membrane permeability ($P$) in the upper and lower plots of Fig. 6, respectively. The known ratios for Lorentzian and Gaussian peaks are indicated in the plots for reference. In Figs. 6(a) and 6(d), lower ratios of the single peaks corresponding to high exchange rates can be seen. This result indicates convergence towards a Gaussian-like distribution of frequencies at high exchange rates. Further discussion of the lineshapes generated by the two models is provided in section 5.

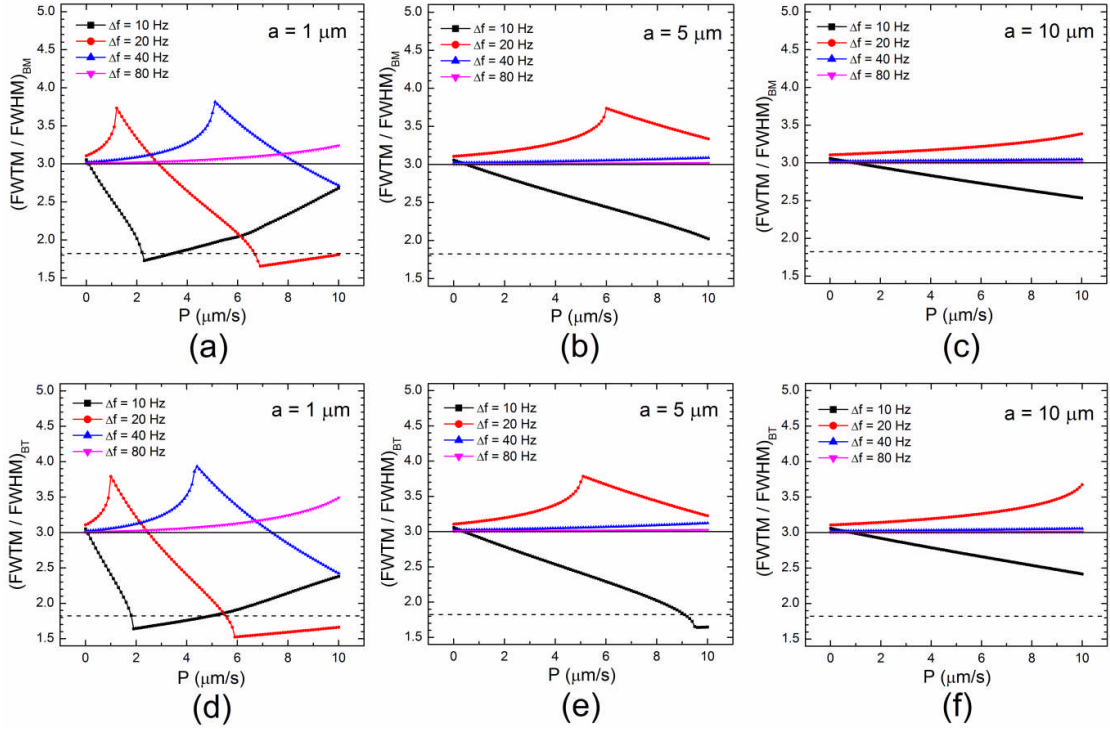

**Fig. 6.** The upper-half panels (a)-(c) show plots of the ratio between the FWTM and FWHM metrics of the theoretical spectra generated using the B-M model $(\text{FWTM}/\text{FWHM})_{\text{BM}}$. The lower-half panels (d)-(f) show the plots of the same ratios corresponding to the B-T model $(\text{FWTM}/\text{FWHM})_{\text{BT}}$. The three plots grouped in each half correspond to three different radii ($a$) and four different chemical shifts ($\Delta f$) for each plot. For reference, the continuous and dashed horizontal lines indicate the ratios corresponding to a Lorentzian peak ($= 3$) and a Gaussian peak ($= \sqrt{\ln 10/\ln 2} \cong 1.82$), respectively.

## 5. Discussion

The results of Fig. 5 from section 4 show that the B-M and B-T models predict the same broadening of the spectra in the limit of increasing sizes ($a$), increasing chemical shifts ($\Delta f$), and decreasing membrane permeability ($P$). This is equivalent to stating that the spectral differences between the two models reduce progressively for higher exchange times ($\tau_e$) and large chemical shifts ($\Delta f$). This result stems from the basic assumptions of the two models. Essentially, the B-M model describes the diffusive exchange process using an average exchange time ($\bar{\tau}_e$), while the B-T model implicitly involves a distribution of exchange times $P(\tau_e)$. A graphical description of this explanation is shown in Fig. 7. The distribution of exchange times was assumed to be Gaussian based on the fact that the diffusive exchange times result from the sum of random-walk histories of many particles. Higher exchange times ($\tau_e$) and larger chemical shifts ($\Delta \omega$) correspond to broader distributions $P(\Delta \varphi)$ of accumulated phase difference $\Delta \varphi$ in time $\tau_e$. In these cases, the contributions of the high and low exchange times to the total transverse magnetization become negligible due to cancelations of exchanged microscopic transverse magnetizations with opposite phases. These cancellations qualitatively explain the quantitative results from Fig. 5.

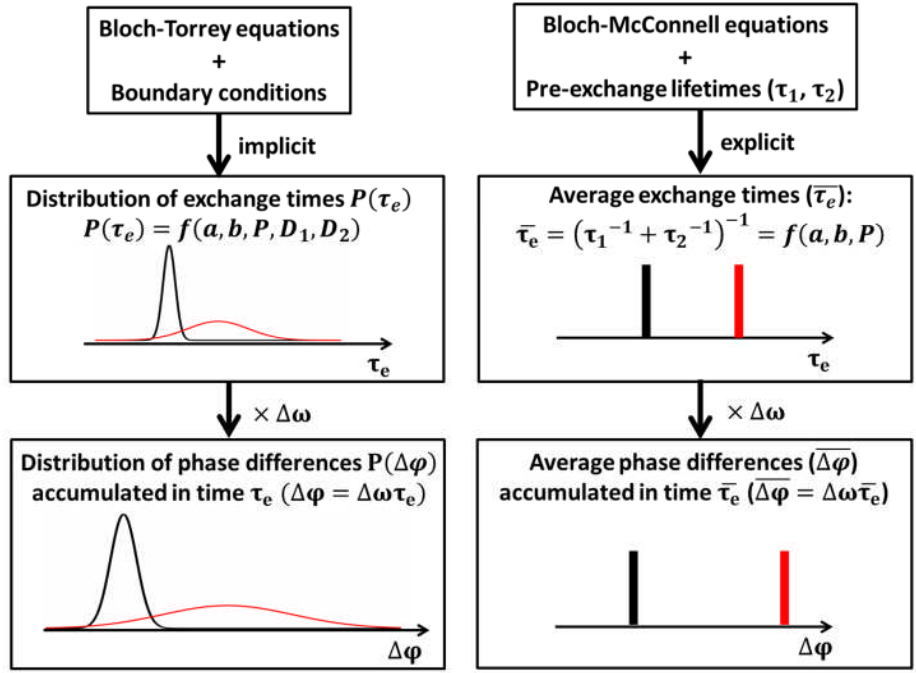

**Fig. 7.** Graphical representation of the exchange time ($\tau_e$) and accumulated phase ($\Delta\varphi = \Delta\omega\tau_e$) distributions for the B-T model (left-hand panels) and B-M model (right-hand panels). The solution of the Bloch-Torrey equations with boundary conditions problem implies a continuous distribution of exchange times $P(\tau_e)$. This corresponds to a continuous distribution of phase differences $P(\Delta\varphi)$ between the two compartments accumulated in time $\tau_e$. The low and high value distributions are shown in black and red lines, respectively. The high exchange time distributions are broader than the low exchange time counterparts, and correspond to larger volumes (large $a$ and $b$), lower diffusion coefficients ($D_1$ and $D_2$), and lower membrane permeability ($P$). The phase difference distributions $P(\Delta\varphi)$ can be calculated from the exchange time distribution $P(\tau_e)$ using a change of variable transformation: $P(\Delta\varphi)d(\Delta\varphi) = P(\tau_e)d\tau_e$. In the right-hand panels the continuous distributions are replaced by single average values.

The differences between the spectral broadening metrics (FWHM and FWTM) are noticeably nonlinear and, in principle, cannot be extrapolated with high accuracy to other physical systems. However, the equal spectral broadening predicted by the two models for high chemical shifts ($\Delta f$) on the scale of the exchange rate ($\tau_e^{-1}$), is also valid for diffusive exchange systems with a different geometrical configuration. Mathematically, this condition is equivalent to:

$$\Delta f/\tau_e^{-1} > 1. \tag{20}$$

The exchange rate ($\tau_e^{-1}$) can be evaluated using the pre-exchange lifetimes method described in appendix A and the chemical shift ($\Delta f$) is readily available from the experimentally acquired NMR spectra. The condition expressed by Eq. (20) is fulfilled in many applications given that many modern NMR spectrometers use a high static magnetic field ($B_0 > 1$ T).

The differences between the spectral broadening predicted by the two models in the case $\Delta f / \tau_e^{-1} \lesssim 1$ can be explained using the qualitative arguments from the first paragraph of this section. If the scale of the exchange rates ($\tau_e^{-1}$) is larger than the chemical shift ($\Delta f$), the rate of microscopic transverse magnetizations cancellations due to opposite phases is reduced explaining, in part, the broadening differences seen in Fig. 5. It is not evident, however, if these differences originate entirely from using different transverse relaxation equations or, partially, from the lifetimes calculations provided in Appendix A. In Figs. 5(a), (d), and (e), the second inflection point marks the full coalescing of the two resonant peaks into a single peak. In this regime ($\Delta f / \tau_e^{-1} \ll 1$), the B-M model consistently predicts a larger broadening of the spectrum than the B-T model, suggesting that the slow-exchange spins contributions to the transverse relaxation process dominate the fast-exchange spin contributions in the latter model. This indicates that fast diffusive exchange systems (relative to the chemical shift $\Delta f$) are more realistically modeled using the B-T equations.

The diffusive exchange is a kinetic process independent of the static magnetic field. However its strength $B_0$ determines indirectly exchange broadening of the NMR lineshapes. Hence, the condition from Eq. (20) does not necessarily indicate a "slow exchange" regime. The criteria used to define the three categories of exchange regimes (slow, fast, and intermediate) vary in the NMR literature depending on the selection of rate scales of the physical processes. The misuse of the approximate relationships corresponding to these regimes can lead to inaccuracies in the data analysis of NMR spectra as demonstrated in the study from reference [15].

The diffusion coefficients ($D_1$ and $D_2$) were not varied to obtain the results from section 4. The variation of diffusion coefficients in the range $(10^{-9} - 10^{-8})$ m$^2$ s$^{-1}$ does not significantly change the broadening of the spectra as shown in Fig. 8. On the other hand, the variation of the membrane permeability ($P$) in the range $(10^{-2} - 10)$ μm s$^{-1}$ significantly increases the magnitude of the diffusive exchange process as indicated by the plots from Figs. (3) - (5). The observed difference can be explained using a time scale analysis. The average lifetime $\tau$ of a particle in a sphere of radius $a$ and diffusion coefficient $D$ can be calculated using two different equations. The first calculation uses Eq. (3) and it takes into account the presence of the permeable membrane. The second equation gives the average lifetime of a diffusing particle within the sphere [17]:

$$\tau = \frac{4a^2}{15D}. \tag{21}$$

For a constant radius $a = 1$ μm, a variation of the membrane permeability from $P_1 = 1$ μm/s to $P_2 = 10$ μm/s, corresponds to a lifetime variation $\Delta \tau \cong 0.3$ s using Eq. (3). For the same radius, a variation of the diffusion coefficient from $D_1 = 10^{-9}$ m$^2$ s$^{-1}$ to $D_2 = 10^{-8}$ m$^2$ s$^{-1}$, gives only a lifetime variation $\Delta\tau = 0.24$ ms using Eq. (21). The time scale ratio of about three orders of magnitude indicates that the rate of the diffusive exchange process is primarily determined by the membrane permeability and the compartment size. On the scale selected for this study, the diffusion coefficient has a small contribution.

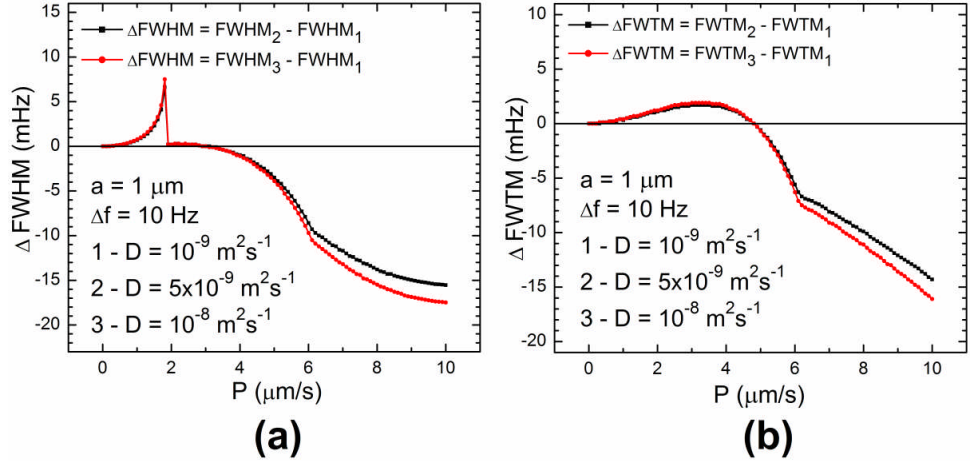

**Fig. 8.** Plots of the FWHM (a) and FWTM (b) differences corresponding to three diffusion coefficients ($D$). The diffusion coefficients are set to be equal in the two compartments ($D_1 = D_2 = D$). The three diffusion coefficients ($D$), inner radius ($a$), and chemical shift ($\Delta f$) values are indicated in the lower-left corner of each plot. Small differences of up to 20 mHz can be noticed for the two spectral broadening metrics. The discontinuity from plot (a) is due to the coalescing peaks and it is the equivalent of the slope discontinuity point from Fig. 4(a).

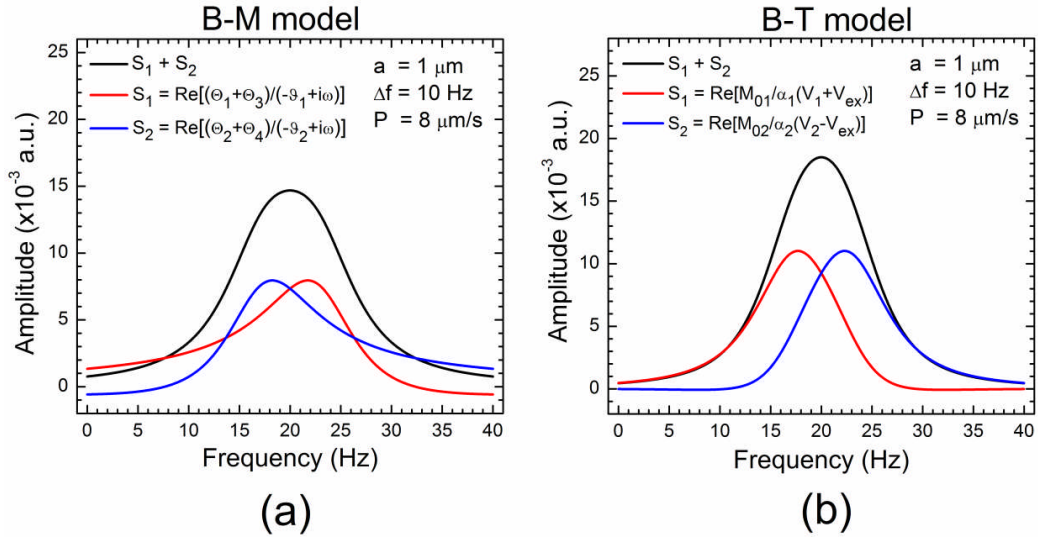

**Fig. 9.** Two single-peak spectra (black lines) corresponding to the B-M model (a) and B-T model (b). Each single-peak spectrum is the sum of the two distinct lineshapes indicated by the red and blue lines. The resonant frequencies are 15 Hz and 25 Hz and the diffusion coefficients for the B-T model are $D_1 = D_2 = 5 \times 10^{-9}$ m$^2$/s. Three other physical parameters ($a$, $\Delta f$, and $P$) are indicated in the upper-right corner of each plot.

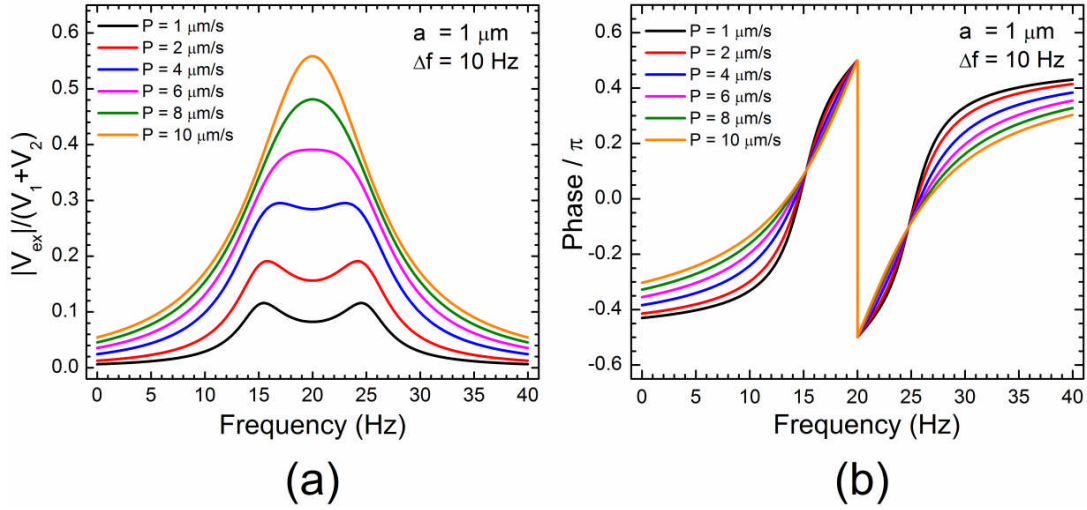

**Fig. 10.** Plots of the relative magnitude (a) and phase (b) of the complex volumetric function $V_{ex}$ from the B-T model for six different values of the membrane permeability indicated in the upper-left corner. The phase value at 20 Hz is $\pi/2$ in plot (b). The resonant frequencies are 15 Hz and 25 Hz and the diffusion coefficients for the B-T model are $D_1 = D_2 = 5 \times 10^{-9}$ m$^2$/s. The physical parameters $a$ and $\Delta f$ are indicated in the upper-right corner of each plot.

The plots from Fig. 6 indicate a similar trend of the lineshapes generated by the two models for different membrane permeability values. For increasing exchange rates the single-peak spectra converge towards Gaussian-like shapes. An example of these single-peak spectra for the two models is shown in Fig. 9. The red and blue lines of the B-M model (Fig. 9a) correspond to the two terms from the right-side of Eq. (9) and show the real part of two phase-shifted complex-valued Lorentzian functions centered on frequencies $\Im(\vartheta_1)$ and $\Im(\vartheta_2)$, respectively. The red and blue lines of the B-T model (Fig. 9b) correspond to the real parts of the two terms from the right-side of Eq. (15). If the notations from section 2.2 are used, the functions $\Re(M_{01}/\alpha_1)$ and $\Re(M_{02}/\alpha_2)$ are standard Lorentzian functions describing the non-exchange NMR peaks. The frequency-domain complex volumetric function $V_{ex}$ from Eq. (15) has a complicated analytical form and, therefore, is more difficult to interpret. This function contains all the diffusive exchange parameters contained within the B-T model. The relative magnitude ($V_{ex}/(V_1 + V_2)$) and phase ($\tan^{-1}[\Im(V_{ex})/\Re(V_{ex})]/\pi$) of $V_{ex}$ are plotted in Fig. 10 for six values of the membrane permeability. According to the convolution theorem, the frequency-domain product between the functions $V_{ex}$ and $(M_{01}/\alpha_1 - M_{02}/\alpha_2)$ from Eq. (15) becomes the time-domain convolution of their inverse Fourier transforms. The frequency-domain function $|V_{ex}|$ takes a Gaussian shape for high exchange rate values as seen in Fig. 10(a). Its inverse Fourier transform $\mathcal{F}^{-1}(V_{ex})$ has also a Gaussian shape in the time-domain and is consistent with the assumption that the B-T model implies a normal distribution of the exchange times. The physical interpretation of the complex function $\mathcal{F}^{-1}(V_{ex})$ in the time-domain requires further analysis in a future study.

## 6. Conclusions

Theoretical NMR spectra were numerically generated from the exact solutions of the B-M and B-T equations which are commonly employed in modelling diffusive exchange systems. The spectral broadening was measured using the FWHM and FWTM metrics which were numerically calculated. The results indicated equality of the spectral broadening in the limit of large chemical shifts relative to the exchange rate ($\Delta f/\tau_e^{-1} > 1$) and differences in all other cases ($\Delta f/\tau_e^{-1} \lesssim 1$). Therefore, systems characterized by high exchange rate on the chemical shift scale ($\Delta f/\tau_e^{-1} < 1$) are better modeled using realistic diffusive exchange models.

**Acknowledgements**

M.R.G. acknowledges financial support from the Mount Allison University, the National Science and Engineering Research Council (NSERC), and the Canadian Research Chair (CRC) program.

# Appendix A

The average lifetime of a particle in a compartment of volume $V$ and area $A$ bounded by a membrane of diffusive permeability $P$ can be derived using the pre-exchange lifetimes method illustrated in Fig. A1.

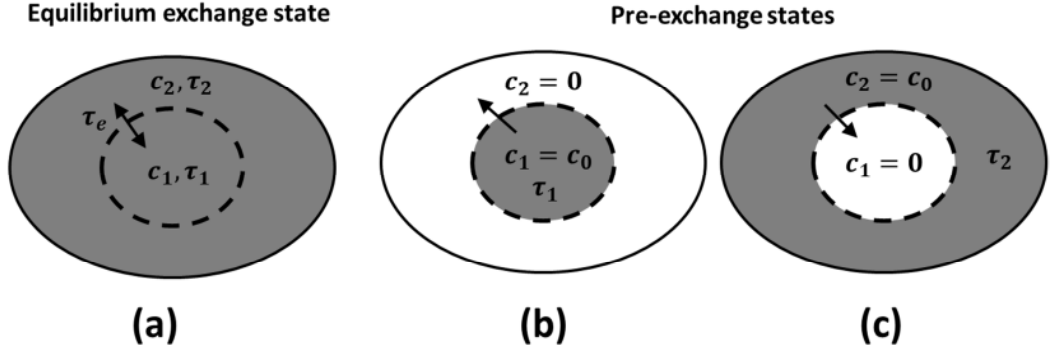

**Fig. A1.** Pictorial description of the pre-exchange method to calculate the average lifetimes $\tau_1$ and $\tau_2$. The physical system consists of two compartments separated by a permeable thin membrane (dashed line). In the equilibrium exchange state (a) the particles exchange compartments through the membrane such that the net flux of particles is zero. In the pre-exchange states (b) and (c) one compartment is assumed to be a particle sink with null concentration.

It is assumed that the thickness of the membrane is much smaller than the dimensions of the compartment (i.e. thin membrane). Thus, the inner side area of the membrane denoted with $A$ is equal to the outer side area. In the pre-exchange states, one of the compartments is a particle sink (null concentration $c = 0$) and the other has an initial volumetric concentration $c(t = 0) = c_0$. In these states, particles will flow through the permeable membrane from the compartment of concentration $c_0$ towards the compartment acting as a particle sink. Conservation of the number of particles requires that the variation of the number of particles within the compartment $dN_{in}$ is balanced by the variation of the number of particles leaving the compartment $dN_{out}$ in the infinitesimal time interval $dt$:

$$dN_{in} + dN_{out} = 0. \tag{A1}$$

The rate of particles leaving the compartment ($dN_{out}/dt$) through the membrane of permeability $P$ and area $A$ is given by:

$$\frac{dN_{out}}{dt} = AP\Delta c. \tag{A2}$$

The quantity $\Delta c$ denotes the concentration difference across the membrane. In the pre-exchange states depicted in Fig. A1(b) and Fig.A1(c) this quantity is equal to the concentration $c(t)$ of particles left after a time interval $t$. In this case, Eq. (A2) becomes:

$$\frac{dN_{out}}{dt} = APc(t). \tag{A3}$$

The rate of the number of particles left in the compartment ($dN_{in}/dt$) at time $t$ is equal to:

$$\frac{dN_{in}}{dt} = V\frac{dc}{dt}, \tag{A4}$$

where the $dc/dt$ represents the particle concentration rate within the compartment at time $t$. Combining Eqs. (A1), (A3) and (A4) the following equation is derived:

$$-\frac{1}{c}\frac{dc}{dt} = P\frac{A}{V}. \tag{A5}$$

The significance of the quantity $P(A/V)$ is evident if Eq. (A5) is integrated with initial condition $c(t=0) = c_0$. The following time dependence of the concentration $c(t)$ is obtained:

$$c(t) = c_0 e^{-P\frac{A}{V}t}. \tag{A6}$$

It follows that the exponential time decay of the particle concentration $c(t)$ has a time constant $\tau$ which can be interpreted as the average lifetime of the particles within the compartment and is given by:

$$\tau^{-1} = P\frac{A}{V}. \tag{A7}$$

It is noteworthy that the average lifetime $\tau$ depends only on the membrane permeability $P$ and area-to-volume ratio $A/V$, and is independent on the particle diffusivity in the two compartments. This result is a consequence of the assumption that the volumetric concentration of particles within the compartment is homogenous. Applying Eq. (A7) to the pre-exchange states of the two compartments, one obtains the following results for the average lifetimes $\tau_1$ and $\tau_2$:

$$\tau_1^{-1} = P\frac{A}{V_1}, \tag{A8}$$

$$\tau_2^{-1} = P\frac{A}{V_2}. \tag{A9}$$

In the exchange equilibrium state shown in Fig.A1(a), the volumetric concentrations in the two compartments $c_1$ and $c_2$ are equal:

$$c_1 = c_2 = \frac{N_1}{V_1} = \frac{N_2}{V_2}, \tag{A10}$$

where $N_1$ and $N_2$ are the number of particles in the two compartments at equilibrium.
From Eqs. (A8)-(A10) it follows that:

$$\frac{\tau_1}{\tau_2} = \frac{V_1}{V_2} = \frac{N_1}{N_2} = \frac{N_1/(N_1+N_2)}{N_2/(N_1+N_2)} = \frac{p_1}{p_2}, \tag{A11}$$

where $p_1$ and $p_2$ are the relative particle populations in the two compartments.
The equilibrium exchange process is quantified by the exchange rate $\tau_e^{-1}$ which is defined by:

$$\tau_e^{-1} = \tau_1^{-1} + \tau_2^{-1}. \tag{A12}$$

Following the physical interpretation of the average lifetimes $\tau_1$ and $\tau_2$, the exchange rate $\tau_e^{-1}$ indicates how fast a percentage equal to $(1 - e^{-1}) \times 100 \cong 63\%$ of the total number of particles $N_1 + N_2$ have exchanged compartments once.

## Appendix B

The Fourier Transform of the function $FID(t)$ from Eq. (8) in the main text is noted with $\mathcal{F}[FID(t)]$ and is given by:

$$\mathcal{F}[FID(t)] = (\Theta_1 + \Theta_3) \int_0^\infty e^{(\vartheta_1 - i\omega)t} dt + (\Theta_2 + \Theta_4) \int_0^\infty e^{(\vartheta_2 - i\omega)t} dt. \tag{B1}$$

Further, the right-hand side integrals from Eq.(B1) can be solved to yield:

$$\mathcal{F}[FID(t)] = \frac{\Theta_1 + \Theta_3}{-\vartheta_1 + i\omega} + \frac{\Theta_2 + \Theta_4}{-\vartheta_2 + i\omega} + \lim_{t \to \infty} \left[ \frac{\Theta_1 + \Theta_3}{\vartheta_1 - i\omega} e^{(\vartheta_1 - i\omega)t} + \frac{\Theta_2 + \Theta_4}{\vartheta_2 - i\omega} e^{(\vartheta_2 - i\omega)t} \right]. \tag{B2}$$

The third complex term from the right-hand side of Eq. (B2) is equal to zero in the limit $t \to \infty$ because $\Re(\vartheta_1) < 0$ and $\Re(\vartheta_2) < 0$. These two inequalities follow from the next four equations which can be derived from Eqs. (7c), (7d), (7i), and (7l):

$$\Re(\vartheta_1) = -\frac{1}{2}\left( R_1 + R_2 + \tau_1^{-1} + \tau_2^{-1} - \sqrt{\frac{\sqrt{\psi^2 + \xi^2} + \psi}{2}} \right), \tag{B3a}$$

$$\Re(\vartheta_2) = -\frac{1}{2}\left( R_1 + R_2 + \tau_1^{-1} + \tau_2^{-1} + \sqrt{\frac{\sqrt{\psi^2 + \xi^2} + \psi}{2}} \right), \tag{B3b}$$

where,

$$\psi = (R_1 - R_2 + \tau_1^{-1} - \tau_2^{-1})^2 - (\omega_{01} - \omega_{02})^2 + 4\tau_1^{-1}\tau_2^{-1}, \tag{B3c}$$

$$\xi = -2(\omega_{01} - \omega_{02})(R_1 - R_2 + \tau_1^{-1} - \tau_2^{-1}). \tag{B3d}$$

The spectrum $S(\omega)$ is then equal to the real part of the remaining two terms from the right-hand side of Eq. (B2):

$$S(\omega) = \Re\left( \frac{\Theta_1 + \Theta_3}{-\vartheta_1 + i\omega} + \frac{\Theta_2 + \Theta_4}{-\vartheta_2 + i\omega} \right). \tag{B4}$$

Equation (9) from subsection 2.1 is obtained by multiplying the two fractions from the right-hand side of Eq. (B4) with the complex conjugates of their denominators.

## Appendix C

The solutions of Eqs. (13a) and (13b) with boundary conditions expressed in Eqs. (14a)-(14d) are given by the following equations:

$$\widetilde{M}_1(r, s) = \frac{\mu_1}{\xi_1^2} + q_1 u_1(r, s), \tag{C1a}$$

$$\widetilde{M}_2(r, s) = \frac{\mu_2}{\xi_2^2} + q_2 u_2(r, s), \tag{C1b}$$

where,

$$\mu_j = \frac{M_{0j}}{D_j}, \quad j = \{1, 2\}, \tag{C2a}$$

$$\xi_j = \left( \frac{\alpha_j}{D_j} \right)^{1/2}, \quad j = \{1, 2\}, \tag{C2b}$$

$$\alpha_j = s + i\omega_{0j} + R_j = R_j - i(\omega - \omega_{0j}), \quad j = \{1, 2\}, \tag{C2c}$$

$$q_1 = \frac{\lambda\left(\frac{\mu_1}{\xi_2{}^2} - \frac{\mu_2}{\xi_1{}^2}\right)}{\frac{D_2}{P}\frac{\partial u_2}{\partial r}(a,s) - u_2(a) + \lambda u_1(a,s)},\qquad(C2d)$$

$$q_2 = \frac{\frac{\mu_2}{\xi_2{}^2} - \frac{\mu_2}{\xi_1{}^2}}{\frac{D_2}{P}\frac{\partial u_2}{\partial r}(a,s) - u_2(a,s) + \lambda u_1(a,s)},\qquad(C2e)$$

$$u_1(r,s) = \frac{\sinh(\xi_1 r)}{\xi_1 r},\qquad(C2f)$$

$$u_2(r,s) = \frac{\xi_2 b\cosh[\xi_2(r-b)] + \sinh[\xi_2(r-b)]}{\xi_2 r},\qquad(C2g)$$

$$\lambda = \frac{D_1 \frac{\partial u_2}{\partial r}(a)}{D_2 \frac{\partial u_1}{\partial r}(a)}.\qquad(C2h)$$

The complex signal $S(s)$ is found by integrating the complex magnetizations $\widetilde{M}_1(r,s)$ and $\widetilde{M}_2(r,s)$ over the compartment volumes $V_1$ and $V_2$ in spherical polar coordinates:

$$S(s) = 4\pi\left[\int_0^a r^2 \widetilde{M}_1(r,s)dr + \int_a^b r^2 \widetilde{M}_2(r,s)dr\right].\qquad(C3)$$

Using Eqs. (C1a) and (C1b), Eq. (C3) can be simplified to the following form:

$$S(s) = \frac{\mu_1}{\xi_1{}^2} V_1 + \frac{\mu_2}{\xi_2{}^2} V_2 + 4\pi(q_1 I_1 + q_2 I_2).\qquad(C4)$$

In Eq. (C3) the following notations were used:

$$I_1 = \int_0^a r^2 u_1(r,s) dr,\qquad(C5a)$$

$$I_2 = \int_a^b r^2 u_2(r,s) dr.\qquad(C5b)$$

Integrals $I_1$ and $I_2$ can be readily solved using integration by parts:

$$I_1 = \frac{a^2 \frac{\partial u_1}{\partial r}(a,s)}{\xi_1{}^2},\qquad(C6a)$$

$$I_2 = -\frac{a^2 \frac{\partial u_2}{\partial r}(a,s)}{\xi_2{}^2}.\qquad(C6b)$$

Using the results of Eqs. (C6a) and (C6b) along with the expressions of $q_1$ and $q_2$ from Eqs. (C2e) and (C2f) it can be demonstrated after some algebraic manipulations that:

$$q_1 I_1 + q_2 I_2 = \frac{Pa^2\left(\frac{M_{01}}{\alpha_1} - \frac{M_{02}}{\alpha_2}\right)\left(\frac{1}{\alpha_1} - \frac{1}{\alpha_2}\right)}{P\left[\frac{u_2(a,s)}{D_2\frac{\partial u_2}{\partial r}(a,s)} - \frac{u_1(a,s)}{D_1\frac{\partial u_1}{\partial r}(a,s)}\right] - 1}.\qquad(C7)$$

Going back to the definitions of $u_1(r,s)$ and $u_2(r,s)$ functions from Eqs. (C2f) and (C2g), the first term of the denominator from Eq. (C7) can be calculated. The result is:

$$P\left[\frac{u_2(a,s)}{D_2 \frac{\partial u_2}{\partial r}(a,s)} - \frac{u_1(a,s)}{D_1 \frac{\partial u_1}{\partial r}(a,s)}\right] = Pa\left[\frac{F_2}{D_2} - \frac{F_1}{D_1}\right],\qquad(C8)$$

where,

$$F_1(\omega_{01}, R_1, D_1, a, b) = \frac{\tanh(\xi_1 a)}{\xi_1 a - \tanh(\xi_1 a)}, \tag{C9a}$$

$$F_2(\omega_{02}, R_2, D_2, a, b) = \frac{\xi_2 b - \tanh[\xi_2(b-a)]}{(1-\xi_2^2 ab)\tanh[\xi_2(b-a)] - \xi_2(b-a)} \tag{C9b}$$

Using the results from Eqs. (C7) and (C8), the complex signal from Eq. (C4) can be written as:

$$S(s) = \frac{M_{01}}{\alpha_1}(V_1 + V_{ex}) + \frac{M_{02}}{\alpha_2}(V_2 - V_{ex}), \tag{C10}$$

with,

$$V_{ex} = -\frac{4\pi a^2 P(\alpha_1^{-1} - \alpha_2^{-1})}{Pa\left(\frac{F_2}{D_2} - \frac{F_1}{D_1}\right) - 1}. \tag{C11}$$

**References**

[1] H.S. Gutowsky, C.H. Holm, Rate processes and nuclear magnetic resonance spectra. II. Hindered internal rotation of amides, J. Chem. Phys. 25 (1956) 1228-1234.

[2] A.D. Bain, Chemical exchange in NMR, Progr. Nucl. Magn. Reson. Spectrosc. 43 (2003) 63-103.

[3] D.E. Woessner, Brownian motion and its effects in NMR chemical exchange and relaxation in liquids, Conc. Magn. Reson. 8 (1995) 397-421.

[4] K.R. Brownstein, C.E. Tarr, Importance of classical diffusion in NMR studies of water in biological cells, Phys. Rev. A 19 (1979) 2446-2453.

[5] G.E. Santyr, I. Kay, R.M. Henkelman, M.J. Bronskill, Diffusive exchange analysis of two-component $T_2$ relaxation of red-blood-cell suspensions containing Gadolinium, J. Magn. Reson. 90 (1990) 500-513.

[6] C.S. Landis, X. Li, F.W. Telang, P.E. Molina, I. Palyka, G. Vétek, C.S. Springer Jr., Equilibrium transcytolemmal water-exchange kinetics in skeletal muscle in vivo, Mag. Reson. Med. 42 (1999) 467-478.

[7] J.D. Quirk, G.L. Bretthorst, T.Q. Duong, A.Z. Snyder, C.S. Springer Jr., J.J.H. Ackerman, J.J. Neil, Equilibrium water exchange between the intra- and extracellular spaces of mammalian brain, Magn. Reson. Med. 50 (2003) 493-499.

[8] S. Michaeli, H. Gröhn, D.J. Sorce, R. Kauppinen, C.S. Springer Jr., K. Uğurbil, M. Garwood, Exchange-influenced $T_{2\rho}$ contrast in human brain images measured with adiabatic radio frequency pulses, Magn. Reson. Med. 53 (2005) 823-829.

[9] H.M. McConnell, Reaction rates by nuclear magnetic resonance, J. Chem. Phys. 28 (1958) 430-431.

[10] H.C. Torrey, Bloch equations with diffusive terms, Phys. Rev. 104 (1956) 563-565.

[11] F. Bloch, Nuclear induction, Phys. Rev. 70 (1946) 460-474.

[12] J. Schotland, J.S. Leigh Jr., Exact solutions of the Bloch equations with *n*-site chemical exchange, J. Magn. Reson. 51 (1969) 48-55.

[13] J.S. Leigh Jr., Relaxation times in systems with chemical exchange: some exact solutions, J. Magn. Reson. 4 (1971) 308-311.

[14] A.C. McLaughlin, J.S. Leigh Jr., Relaxation times in systems with chemical exchange: approximate solutions for the nondilute case, J. Magn. Reson. 9 (1973) 296-304.

[15] D.F. Hansen, J.J. Led, Implications of using approximate Bloch-McConnell equations in NMR analyses of chemically exchanging systems: application to the electron self-exchange of plastocyanin, J. Magn. Reson. 163 (2003) 215-227.

[16] N.-M. Chao, S.H. Young, M.-M. Poo, Localization of cell membrane components by surface diffusion in a "trap", Biophys. J. 36 (1981) 139-153.

[17] M.J. Hey, F. Al-Sagheer, Interphase transfer rates in emulsion studied by NMR spectroscopy, Langmuir 10 (1994) 1370-1376.

[18] J.V. Sehy, A.A. Banks, J.J.H. Ackerman, J.J. Neil, Importance of intracellular water apparent diffusion to the measurement of membrane permeability, Biophys. J. 83 (2002) 2856-2863.

[19] S.-T. Chen, C.S. Springer Jr., Ionophore-catalyzed cation transport between phospholipid inverted micelles manifest in DNMR, Biophys. Chem. 14 (1981) 375-388.

[20] P.S. Belton, B.P. Hills, The effects of diffusive exchange in heterogeneous systems on NMR lineshapes and relaxation processes, Mol. Phys. 61 (1987) 999-1018.

[21] P.S. Belton, B.P. Hills, The effects of exchange and interfacial reaction in two-phase systems on NMR lineshapes relaxation processes, Mol. Phys. 65 (1988) 313-326.

[22] M.R. Gherase, J.C. Wallace, A.R. Cross, G.E. Santyr, Two-compartment radial diffusive exchange analysis of the NMR lineshape of $^{129}$Xe dissolved in a perfluorooctyl bromide solution, J. Chem. Phys. 125 (2006) 044906

[23] C. Maurel, Aquaporins and water permeability of plant membranes, Annu. Rev. Plant Physiol. Plant Mol. Biol. 48 (1997) 399-429.

[24] D. Le Bihan, R. Turner, P. Douek, N. Patronas, Diffusion MR imaging: clinical applications, Am. J. Roentgenol. 159 (1992) 591-599.